\input epsf
\documentstyle[11pt,epsf]{article}
\title{\bf{Count Models Based on Weibull Interarrival Times}}

\author{Moshe Adrian, Eric T. Bradlow, Peter S. Fader, and Blake McShane\thanks{Moshe Adrian is a Postdoctoral Research
Assistant Professor
in Mathematics at the University of Utah, Eric T. Bradlow is an Associate Professor of Marketing and Statistics, Peter S.
Fader is the Frances and Pei-Yuan Chia Professor, Professor of Marketing, and Blake McShane is an Assistant Professor
of Marketing at Northwestern University.  All correspondence on this manuscript should
be sent to Eric T. Bradlow, The Wharton School of the University of Pennsylvania, 3730 Walnut Street, Philadelphia, PA 19104,
ebradlow@wharton.upenn.edu, (215) 898-8255.  The authors thank Rainer Winkelmann for useful suggestions and comments and for sending
us his data.}}

\date{}

\renewcommand{\baselinestretch}{1.2}\large\normalsize
\begin{document}
\maketitle
\begin{abstract}

The widespread popularity and use of both the Poisson and Negative-Binomial models for count data arises, in part,
from their derivation as the number of arrivals in a given time period assuming exponenitally distributed interarrival times
(without and with heterogeneity in the underlying base rates respectively).  However, with that {\em clean} theory comes some
limitations including limited flexibility in the assumed underlying arrival rate distribution and
the inability to model underdispersed counts (variance less than the mean).  While extant research has addressed some
of these issues, it has self-proclaimed itself to be ``not ideal and somewhat inadequate''.

In this research, we present a model that extends and responds to the {\em call} from Winkelmann (1995), yet due
to computational tractability was previously thought to be infeasible.  In particular, we
introduce here a generalized model for count data based upon an assumed Weibull interarrival process that nests the Poisson
and Negative-Binomial models as special cases.  The computational intractability is overcome by deriving the Weibull
count model using a polynomial expansion which then allows for closed-form inference (integration term-by-term)
when incorporating heterogeneity due to the conjugacy of the expansion and a commonly employed Gamma distribution.

In addition, we demonstrate that this new Weibull counting model can: (a) sometimes alleviate the need for heterogeneity suggesting
that what many think is overdispersion may just be model misfit due to a different and more flexible timing model (Weibull
versus exponential), (b) model both over and under dispersed count data, (c) allow covariates to be introduced
straightforwardly through the hazard function, and (d) be computed in standard software.  In fact, we demonstrate the efficacy
of our approach using a worked-out example run in Microsoft Excel.

\end{abstract}

\newpage

\section{Introduction}\label{intro}

\renewcommand{\baselinestretch}{1.8}\large\normalsize

\begin{itemize}

\item{{\em Quote 1: Although the Weibull distribution is preferred in duration analysis for its closed-form hazard function, ....}}

\item{{\em Quote 2: For noninteger $\alpha$, no closed-form expression is available for the Gamma (counting) model ...}}

\item{{\em Quote 3: The regression is for waiting times and not directly for counts ... hence the estimated parameters have to be
interpreted (carefully and) accordingly.}}

\item[]{Winkelmann (1995)}

\end{itemize}

The ubiquitity of count data, and the use of models for analyzing it, makes (by definition) significant advances in their
analysis a ``big deal''.  The above quotes, taken from Winkelmann's (1995) paper and echoed in Winkelmann (2003), suggests the need
for research into count data models well beyond that which is currently available.

In particular, quote 1 suggests and calls for the development of a
count model based upon a flexible family of arrival curves (e.g.
the Weibull), in contrast to the commonly employed Poisson and
Negative-Binomial models that are based upon a more restrictive
underlying exponential timing process.  As is well-known, if for
no other reason (although as we show there are many more), the
constant (decreasing) hazard function of the exponential
(exponential with gamma heterogeneity) leads to a counting model
where the variance is equal (greater than or equal to) the mean
(Barlow and Proschan, 1965). That is, underdispersed count data
(Cameron and Johansson, 1997; Trivedi and Deb, 1997, Cameron and
Trivedi, 1996, King, 1989) that has been shown to exist in many
domains is entirely left to ``fend for itself''.  On the other
hand, if derivable, the Weibull counting model which allows for
decreasing, constant, or increasing hazard (as the shape parameter
is greater than, equal to, or less than 1 respectively), does not
suffer this limitation.

However, by no means has this problem gone unrecognized and has been addressed, in part, as per quote 2.  In Winkelmann (1995),
they derive a flexible model for count data that allows for both under and over dispersion, based upon a gamma timing model; however,
as stated, this model only has a closed-form solution for integer values of $\alpha$ (the shape parameter), and furthermore,
covariates can not be brought into the model in a ``natural way'' (quote 3) and hence have to be addressed with great care.
Another type of distribution, which is similar to the negative binomial, is the continuous parameter binomial (CPB).  This is
also used to model underdispersed data.  It simplifies to the Poisson when $\alpha = 1$, however it also imposes a theoretical upper
limit on the count variable.  In King (1989) they offered a "generalized event count" (GEC)
distribution, which can handle overdispersion, underdispersion, or
when the mean equals the variance.  It nests the CPB, however has
a similar restriction as the CPB on the count variable (Winkelmann 1995) as well as other features (described below) that can
be improved upon.

In summary, these ``post-publication'' quotes allow us to look back at the problem of modeling count data from a fresh
perspective and ask ourselves ``If one were to try and imagine what an "optimal" model for count data would look like, what
characteristics would that model have?''  Furthermore, where do current existing models fall short and provide an opportunity
for a new model that meets these shortcomings?  We believe that the following ``wish list'' provides a set of reasonable properties:

\begin{itemize}
\item[(1)]{The model should be able to generalize (nest) the most commonly used extant models such as the
Poisson and the negative binomial distribution (NBD) as special cases; thus, when a more simple mathematical story is sufficient, the
model will indicate this.}

\item[(2)]{It should be able to handle both overdispersed and underdispersed data, both of which are likely to be seen in practice.}

\item[(3)]{There should be an underlying story about interarrival times from which the model is derivable.
That is, as in the Poisson and Negative Binomial models, the adoption of the model will be based (in part) on the underlying
behavioral process story that underlies its foundation.}

\item[(4)]{It should be computationally feasible to work with, where if possible a closed-form solution found.}

\item[(5)]{The model should allow for the incorporation of person-level heterogeneity reflecting the fact that individual's
interarrival rates may vary across the inferential population.}

\item[(6)]{Covariate effects should be incorporatable as many (if not most) data sets have explanatory variables.}

\end{itemize}

In this paper, we derive a new model for count data that satisfies
these six criteria in the following ways.  First, our counting
model is based upon an assumed Weibull interarrival time model
that nests the exponential interarrival time model (kernel for the
Poisson and Negative Binomial) when the shape parameter is equal
to 1.  Second, we demonstrate that the Weibull counting model, via
the shape parameter being less than, equal to, or greater than 1
can model overdispersed, equally mean-variance dispersed, and
underdispersed data respectively.  Third, the Weibull interarrival
time story is more rich than the exponential story allowing for
hazard rates that are non-constant.  Fourth, and is one of the
main contributions of this research, according to Winkelmann
(1995) the Weibull counting model is preferred but ``infeasible''
due to its non-closed form nature.  Here we ``solve'' the
previously unsolved non-closed form nature of the Weibull counting
model (see Bradlow, Hardie, and Fader (2002), Everson and Bradlow
(2002), Miller, Bradlow and Dayartna (2004) for similar solutions
for the Negative Binomial, Beta-Binomial and binary logit models
respectively) problem by deriving it using a polynomial expansion
which can indeed be expressed in closed-form.  Fifth, and related
to four, once the model can be expressed as a closed-form sum of
poylnomial terms, the gamma distribution provides a conjugate
heterogeneity distribution reflecting the underlying disperson in
rates across individuals.  This is nice in that not only can the
marginal distribution (marginalized over heterogeneity) be derived
in closed-form, but the conjugacy and use of the Gamma
distribution allows us to nest the Negative Binomial as a special
case (i.e. where the Weibull has shape equal to 1).  Finally, as
we demonstrate, a hazard rate formulation can be directly
implemented in our Weibull counting model (with covariates), and
hence as quote 3 above states, covariates are brought into the
model in a natural way.

The remainder of this paper is laid out as follows.  In the next section, we provide a more detailed description of the major
ways in which extant research has extended basic count data models; however as we will show, not at the interarrival time level.
Section \ref{derivation} contains the derivation
of our Weibull counting model focusing specifically on the polynomial approximation that leads to the closed-form benefits.  In Section
\ref{empirical} we re-analyze the data of Winkelmann (1995) and provide a set of results comparing a sequence of nested models the most
complicated of which has an underlying Weibull arrival process, heterogeneous baseline rates, and covariates, where the simplest is
the special case of the Poisson model (i.e. Weibull shape equal to one, no heterogeneity in rates, and all covariate slopes equal to 0.)
Through the sequence of models we fit, we are able to ascertain which aspects of the model are
doing the ``heavy lifting''.  As we demonstrate,
and is an area for future empirical study, once the Weibull counting process is utilized, the need for underlying heterogeneity is
greatly reduced in our example, and may be more generally.  Finally, we provide some concluding remarks and areas for future research in Section \ref{conclusions}.

\section{Prior Related Research}
\label{priorwork}

The way in which this research contributes to the literature on count data is by generalizing the underlying interarrival time model that is used to derive the count data model, by allowing greater flexibility in its' hazard function.  Or, as in Winkelmann (1995), an underlying arrival process may have
non-constant hazard function, and this affects whether the
arrival time distribution displays "duration dependence".  In particular, if we denote the
mean of the interarrival distribution by $\mu$, the variance
$\sigma^2$, and the hazard function by $$h(t) =
\frac{f(t)}{1-F(t)},$$ where f(t) and F(t) are the density and
cumulative probability functions respectively, we say that the distribution has negative duration dependence if
$\frac{dh(t)}{dt} < 0$ and positive duration dependence if
$\frac{dh(t)}{dt} >0 $.  If the hazard function is
monotonic, then $$\frac{dh(t)}{dt} >0 \Rightarrow \sigma/\mu < 1$$
$$\frac{dh(t)}{dt} =0 \Rightarrow \sigma/\mu = 1$$ $$\frac{dh(t)}{dt} <0 \Rightarrow \sigma/\mu >
1.$$ (see Barlow and Proschan 1965, p. 33).  In particular, as we discuss, the Weibull counting model allows for positive, negative or no duration dependence.  However, duration dependence is but one way (explanation) in which people
have extended arrival time models, and we discuss these briefly.

One can also assume that successive events are dependent in that
the probability of an event occurring depends on the
number of events that have occurred previously, as
compared to the arrival time of the most recent event (duration
dependence).  These models are said to display contagion.  They
have been studied  in the literature on accident proneness (Arbous
and Kerrich 1951; Feller 1943).  For more information, one can
reference Gurland (1995) for a contagious discrete-time model that
leads to the negative binomial in which an occurrence increases
and a nonoccurrence decreases the probability of a future
occurrence.  Other models for occurrence dependence have been
developed by Mullahy (1986), and Gourieroux and Visser (1997).
One can also make the assumption that successive events are
independent but the process intensity varies as a function of
time.  This class are denoted nonhomogeneous Poisson processes and
are described in Lawless (1987).  We believe that an area
for future research, would be a comparison of all of these approaches for allowing flexibility in the count data models; albeit here we
focus on the hazard generalization (duration dependence) as described in detail next.

\subsection{A Modeling Framework}\label{modelframe}

Much extant research on count data has been focused on extending the basic Poisson model (denoted here as model [0]) to account (allow)
for hyperdispersion via a non-constant hazard rate.  The basic ways in which hyperdispersion have been accounted for include: (model [1]) adding covariates to the model, (model [2]) incorporating individual-level heterogeneity for the baseline rates,  and
(model [3]) both [1] and [2].  In particular, if we let

\vspace{-0.2in}

\begin{eqnarray}\label{model1}
[X_{it}|\lambda_{i}] & \sim & {\rm Poisson}(\lambda_{i}{\rm exp}(Z_{it}^{\prime}\beta)),
\end{eqnarray}

\noindent a proportional-hazards framework (Cox, 1972), where
$X_{it}$ is a non-negative integer (count) for unit $i=1,\ldots,I$
on its $t=1,\ldots,T_{i}$-th observation, $\lambda_{i}$ is the
baseline rate for unit $i$, $Z_{it}=(Z_{it1},\ldots,Z_{itP})$ is a
vector of time-varying covariates, and
$\beta^{\prime}=(\beta_{1},\ldots,\beta_{P})$ is a vector of
covariate slopes: model [0] is obtained by setting
$\lambda_{i}=\lambda$ for all $i$ and $Z_{it}^{\prime}\beta=0$ (an
intercept only); model [1] is obtained by setting
$\lambda_{i}=\lambda$ for all $i$ (the Poisson Regression Model);
model [2] is obtained by setting $P=1$, $Z_{it}^{\beta}=0$ and
letting $\lambda_{i} \sim f(\lambda_{i}|\theta)$ (when $f$ is the gamma
distribution then model [2] integrated over the distribution of
$\lambda_{i}$ is the Negative Binomial Distribution); and model
[3] is as given in equation (\ref{model1}) where again
$\lambda_{i} \sim f(\lambda_{i}|\theta)$.  Model [3] is also sometimes
referred to as the Neg-Bin II model when $f(\lambda_i|\theta)$ is the gamma distribution. Later in
Section \ref{empirical}, we compare the results of models [0]-[3]
to those derived in this research.

What is of interest to note is that all of these extensions use the Poisson model (with associated exponential interarrival time) as
their kernel.  That is, these extensions to the model have not been done at the core unit of analysis, the underlying arrival time
distribution, but instead to the count model equivalent from an assumed simple arrival time distribution.  What we have done in this
research is to ``stretch'' the flexibility in the arrival time model to account for richer patterns.  In particular, instead, we
assume that the underlying arrival time distribution for $Y_{ik}$, the $k-th$ arrival for unit $i$ is Weibull (proportional
regression) with density given by:

\begin{eqnarray}\label{model2}
f(Y_{ik}=y|\lambda_{i}, \beta, c) & = & \lambda_{i}{\rm exp}(Z_{it}^{\prime}\beta)cy^{c-1}{\rm exp}(-\lambda_{i}exp(Z_{it}^{\prime}\beta)
 y^{c})
\end{eqnarray}

\noindent where as in (\ref{model1}), $\lambda_{i}{\rm exp}(Z_{it}^{\prime}\beta)$ denotes the proportional regression parameterization
of the rate parameter, and $c$ denotes the shape parameter
of the Weibull distribution.   We note, as before, that when $c=1$, equation (\ref{model2}) simplifies to a heterogeneous
exponential arrival time model with covariates that leads to count models [0]-[3] above.

Thus, directly analogous to models [0]-[3] which are based on an exponential interarrival time, our interest
lies in looking at various reduced-form specifications of model (\ref{model2}).  Specifically, we denote as model [4], the Weibull model
without heterogeneity and without covariates (model [0] analog) such that $\lambda_{i} = \lambda$ and $Z_{it}^{\prime}\beta = 0$.
We label model [5] as the Weibull regression model (without heterogeneity) such that $\lambda_{i}=\lambda$.  Model [6] is the model
in which we allow for heterogeneity in baseline rates $\lambda_{i}$ but do not include covariates ($Z_{it}^{\prime}\beta=0$).  Finally,
model [7] is the fully parameterized model given by (\ref{model2}).  All eight of these models will be fit and results compared in
Section \ref{empirical}.  However, as we have noted, the primary distinction between the Weibull-based models [4]-[7], and their
exponential-based analogs [0]-[3] are that the hazard rate can be decreasing, constant, or increasing.  Provided next in Section
\ref{derivation} is the derivation of the basic Weibull counting model (without heterogeneity and covariates) and the Weibull heterogeneous regression model obtained using (\ref{model2}).

\section{Basic Theory and Definitions}
\label{derivation}

Before discussing the Weibull counting model itself, we describe the general framework utilized to derive the model
that is based upon the relationship between interarrival times and their count model equivalent.
Let $Y_i$ be the time from the measurement time origin at which the $i$-th event
occurs.  Let $X(t)$ denote the number of events that have occurred
up until time $t$.  The relationship between interarrival times and
the number of events is $$Y_i \leq t \Leftrightarrow X(t) \geq
i.$$ We can restate this relationship by saying that the amount of
time at which the $i$-th event occurred from the time origin is less
than or equal to $t$ if and only if the number of events that have
occurred by time $t$ is greater than or equal to $i$.

We therefore have the following relationships that allow us to derive
our Weibull counting model $C_{i}(t)$:

\vspace{-0.2in}

\begin{eqnarray}\label{basicmodel}
C_{i}(t) = P(X(t) = i) & = & P(X(t) \geq i) - P(X(t) \geq i+1) \\ \nonumber
 & = & P(Y_i \leq t) - P(Y_{i+1} \leq t).
\end{eqnarray}

\noindent If we let the cumulative density function (cdf) of $Y_i$ be $F_i(t)$, then $C_{i}(t)=P(X(t) = i) = F_i(t) -
F_{i+1}(t)$.  In the case where the measurement time origin (and
thus the counting) process coincides with the occurrence of an
event, then $F_i(t)$ is simply the i-fold convolution of the common interarrival
time distribution which may or may not have a closed-form solution.  Based upon (\ref{basicmodel}),
we derive our Weibull counting model next based upon a polynomial expansion of F(t).

\subsection{Weibull Counting Model}

We derive the basic Weibull count model, model [4] from above,
by assuming that the interarrival times are independent and
identically distributed Weibull with probability density function (pdf) $f(t) = \lambda ct^{c-1}e^{- \lambda t^c}, (c, \lambda \in
R^+$),  and corresponding cdf $F(t) = 1-e^{-\lambda
t^c}$, which simplifies to the exponential model when $c=1$.  As is known, and represents
the flexibility in our model, the hazard function $h(t)$ is equal
to

\vspace{-0.2in}

\begin{eqnarray}\label{hazard}
h(t) & = & \lambda c t^{c-1}
\end{eqnarray}

\noindent which is monotonically increasing for $c > 1$,
monotonically decreasing for $c < 1$, and constant (and equal to
$\lambda$) when $c = 1$.

If a closed-form solution were to be attainable for the Weibull count model, we would need to be able to evaluate
convolutions of the form $\int_0^t F(t-s)f(s)ds$.   However, this integral does not have a
proper solution for the Weibull density.  Thus, our approach is to derive the Weibull counting
model through a Taylor series approximation to the Weibull density.

In particular, the Taylor series approximation obtained by expanding the exponential pieces ($e^{\lambda t^{c}}$) respectively,
for both the cdf and pdf of the Weibull are:

\vspace{-0.2in}

\begin{eqnarray}\label{Ftaylor}
F(t) & = & \sum_{j=1}^{\infty} \frac{(-1)^{j+1} (\lambda t^c)^j}{\Gamma(j+1)}
\end{eqnarray}

\noindent and

\vspace{-0.2in}

\begin{eqnarray}\label{ftaylor}
f(t) & = & \sum_{j=1}^{\infty} \frac{(-1)^{j+1} cj \lambda^j t^{cj-1}}{\Gamma(j+1)}.
\end{eqnarray}

Utilizing, as in (\ref{basicmodel}), that $C_i(t) = F_i(t) - F_{i+1}(t)$, we obtain
the following recursive relationship that we utilize in deriving the Weibull count model:

\vspace{-0.2in}

\begin{eqnarray}\label{derivation1}
C_{i}(t) & = & \int_0^t F_{i-1}(t-s)f(s)ds - \int_0^t F_i(t-s)f(s)ds \\ \nonumber
 & = & \int_0^t C_{i-1}(t-s)f(s)ds.
\end{eqnarray}

Before proceeding to develop the general solution to the problem, we note that
straightforwardly $F_0(t)$ is 1 for all $t$ and $F_1(t) = F(t)$. Therefore, we have $C_0(t) = F_0(t) -
F_1(t) = e^{-\lambda t^c} = \sum_{j=0}^{\infty} \frac{(-1)^j
(\lambda t^c)^j}{\Gamma(j+1)}.$   Using the recursive formula in (\ref{derivation1}), we can
therefore compute $C_1(t)$:

\begin{eqnarray}\label{derivation2}
C_1(t) & = &  \int_0^t C_0(t-s)f(s)ds  \\
& = & \int_0^t (\sum_{j=0}^{\infty} \frac{(-1)^j
(\lambda(t-s)^c)^j}{\Gamma(j+1)})(\sum_{k=1}^{\infty}
\frac{(-1)^{k+1} ck \lambda^k s^{ck-1}}{\Gamma(k+1)}ds^{F1} \\ \nonumber
 & = & \sum_{j=0}^{\infty} \sum_{k=1}^{\infty} \frac{(-1)^j (-1)^{k+1}
(\lambda)^j (\lambda)^k}{\Gamma(j+1) \Gamma(k+1)} \int_0^t
ck(t-s)^{cj} s^{ck-1}ds \\ \nonumber
 & = &  \sum_{j=0}^{\infty}
\sum_{k=1}^{\infty} \frac{(-1)^j (-1)^{k+1} (\lambda)^j
(\lambda)^k}{\Gamma(j+1) \Gamma(k+1)}
\frac{(t)^{cj}(t)^{ck}\Gamma(cj+1)\Gamma(ck+1)}{\Gamma(cj+ck+1)}
\end{eqnarray}

\noindent Then, by using a change of variables $m=j$ and $l=m+k$, we obtain:

\begin{eqnarray*}
& = &  \sum_{l=1}^{\infty} (\sum_{m=0}^{l-1} \frac{(-1)^m
(-1)^{l-m+1} (\lambda)^m
(\lambda)^{l-m}}{\Gamma(m+1)\Gamma(l-m+1)} \frac{(t)^{cm}
(t)^{cl-cm} \Gamma(cm+1)
\Gamma(cl-cm+1)}{\Gamma(cm+cl-cm+1)}) \\ \nonumber
 & = &  \sum_{l=1}^{\infty} \frac{(-1)^{l+1} (\lambda
t^c)^l}{\Gamma(cl+1)} (\sum_{m=0}^{l-1} \frac{\Gamma(cm+1)
\Gamma(cl-cm+1)}{\Gamma(m+1) \Gamma(l-m+1)} \\ \nonumber
 & = & \sum_{l=1}^{\infty} \frac{(-1)^{l+1} (\lambda t^c)^l
\alpha_m^l}{\Gamma(cl+1)}
\end{eqnarray*}

\noindent where $\alpha_m^l = \sum_{m=0}^{l-1} \frac{\Gamma(cm+1) \Gamma(cl-cm+1)}{\Gamma(m+1) \Gamma(l-m+1)}$.

This suggests a general form for $C_i(t)$, namely:
$\sum_{l=i}^{\infty} \frac{(-1)^{l+i} (\lambda t^c)^l
\alpha_l^i}{\Gamma(cl+1)}$  which is confirmed by

\vspace{-0.2in}

\begin{eqnarray}\label{derivation3}
C_{i+1}(t) & = & \int_0^t C_i(t-s)f(s)ds \\ \nonumber
 & = & \int_0^t (\sum_{j=i}^{\infty} \frac{(-1)^{j+i} (\lambda(t-s)^c)^j
\alpha_j^i}{\Gamma(cj+1)})(\sum_{k=1}^{\infty} \frac{(-1)^{k+1} ck
\lambda^k s^{ck-1}}{\Gamma(k+1)}ds \\ \nonumber
 & = & \sum_{j=i}^{\infty} \sum_{k=1}^{\infty} \frac{(-1)^{j+i} (-1)^{k+1} (\lambda)^j
(\lambda)^k \alpha_j^i}{\Gamma(cj+1) \Gamma(k+1)} \int_0^t
ck(t-s)^{cj} s^{ck-1}ds \\ \nonumber
 & = & \sum_{j=i}^{\infty}
\sum_{k=1}^{\infty} \frac{(-1)^{j+i} (-1)^{k+1} (\lambda)^j
(\lambda)^k \alpha_j^i}{\Gamma(cj+1) \Gamma(k+1)}
\frac{(t)^{cj}(t)^{ck}\Gamma(cj+1)\Gamma(ck+1)}{\Gamma(cj+ck+1)} \\ \nonumber
 & = & \sum_{l=i+1}^{\infty} \frac{(-1)^{l+i+1} (\lambda t^c)^l}{\Gamma(cl+1)}
(\sum_{m=i}^{l-1} \alpha_m^i \frac{\Gamma(cl-cm+1)}{\Gamma(l-m+1)})  \\ \nonumber
 & = & \sum_{l=i+1}^{\infty} \frac{(-1)^{l+1} (\lambda t^c)^l
\alpha_l^{i+1}}{\Gamma(cl+1)}
\end{eqnarray}

\noindent where $\alpha_l^{i+1} = \sum_{m=i}^{l-1} \frac{\Gamma(cl-cm+1)}{\Gamma(l-m+1)}.$

Therefore, we have the main result of this paper, the Weibull counting model:

\vspace{-0.2in}

\begin{eqnarray}\label{derivation4}
P(I(t) = i) & = & C_i(t) = \sum_{j=i}^{\infty} \frac{(-1)^{j+i} (\lambda t^c)^j
\alpha_j^i}{\Gamma(cj+1)} \ \ i = 0,1,2,...
\end{eqnarray}

\noindent where $\alpha_j^0 = \frac{\Gamma(cj+1)}{\Gamma(j+1)} \ \ j = 0,1,2,...$ and
$\alpha_j^{i+1} = \sum_{m=i}^{j-1} \alpha_m^i \frac{\Gamma(cj-cm+1)}{\Gamma(j-m+1)}, {\rm for} \ i = 0,1,2,... \ \ {\rm for} \ \ j =
i+1,i+2,i+3,...$.

We note in addition that the expectation of this counting model is

\begin{eqnarray*}
E(I) & = & \sum_{i=1}^{\infty} \sum_{j=i}^{\infty} \frac{i (-1)^{j+i}
(\lambda t^c)^j \alpha_j^i}{\Gamma(cj+1)}
\end{eqnarray*}

\noindent with variance given by

\begin{eqnarray*}
Var(I) & = & E(I^2) - (E(I))^2 \\
& = & \sum_{i=2}^{\infty} \sum_{j=i}^{\infty} \frac{i^2 (-1)^{j+i}
(\lambda t^c)^j \alpha_j^i}{\Gamma(cj+1)} - (\sum_{i=1}^{\infty}
\sum_{j=i}^{\infty} \frac{i (-1)^{j+i} (\lambda t^c)^j
\alpha_j^i}{\Gamma(cj+1)})^2.
\end{eqnarray*}

\subsection{The Benefits of the Weibull Counting Model}

We now revisit the wish list created in Section \ref{intro}, point-by-point, both to describe those aspects that the basic Weibull
counting model (without covariates and without heterogeneity) given in (\ref{derivation4}) provides, and those that require extensions.

\begin{itemize}
\item[(1)]{The model should be able to generalize (nest) the most commonly used extant models such as the
Poisson and the negative binomial distribution as special cases; thus, when a more simple mathematical story is sufficient, the
model will indicate this.}
\end{itemize}

We note therefore that when we set $c=1$ in (\ref{derivation4}), we do in fact get the
Poisson counting model as $P(I(t) = i) = \sum_{j=i}^{\infty}
\frac{(-1)^{j+i} (\lambda)^j \alpha_j^i}{\Gamma(j+1)}$, a standard result. With regards to the negative binomial model, we discuss this
with respect to item [5] in the wish list in which $\lambda$ is replaced by a heterogeneous proportional hazards
framework specification given by $\lambda_{i}{\rm exp}(X_{i}^{\prime}\beta)$.

\begin{itemize}
\item[(2)]{It should be able to handle both overdispersed and underdispersed data, both of which are likely to be seen in practice.}
\end{itemize}

Through extensive simulations, as the result is unavailable in closed-form, we have
verified that for $0 < c < 1$, the Weibull
count distribution function displays overdispersion, whereas for
$c > 1$, underdispersion is displayed.  That is the underlying
interarrival times have a decreasing (increasing) hazard for $0 <
c < 1 \ \ (c > 1$).  Thus, negative duration dependence is
associated with overdispersion, positive duration dependence with
underdispersion (Winkelmann 1995).  No duration dependence leads
to the Poisson distribution with equal mean and variance.

As one demonstration of these findings, Figures
\ref{fig1} and \ref{fig2} display probability histograms for the Weibull and
Poisson counting models with different parameter values.  Both the Weibull and the Poisson were intentionally chosen to have
identical means (mean set to 2); yet their dispersion is quite different.  In
Figure \ref{fig1}, we have the probability histograms for an underdispersed
Weibull with parameters $c = 1.5$ and $\lambda = 2.93$, and a
Poisson with $\lambda = 2$.  The variance of the
Weibull counting model in this case is 0.880.  In Figure \ref{fig2}, we have
the probability histograms for an overdispersed Weibull with
parameters $c = .5$ and $\lambda = 1.39$, and again the Poisson with
$\lambda = 2$.  The variance of the Weibull counting
model in this case is 3.40, greater than the mean, as expected.

\begin{center}
{\bf Insert Figures \ref{fig1} and \ref{fig2} here}
\end{center}

\begin{itemize}
\item[(3)]{There should be an underlying story about interarrival times from which the model is derivable.
That is, as in the Poisson and Negative Binomial models, the adoption of the model will be based (in part) on the underlying
behavioral process story that underlies its foundation.}
\end{itemize}

The way in which the Weibull interarrival time model generalizes the exponential model to allow for both increasing
and decreasing hazard, as demonstrated in (\ref{hazard}), and its translation into the equivalent count model as
demonstrated in Figures \ref{fig1} and \ref{fig2}, suggests a very flexible arrival time story.

\begin{itemize}
\item[(4)]{It should be computationally feasible to work with, where if possible a closed-form solution found.}
\end{itemize}

\noindent This item, of course, is the entire motivation for this paper and our solution is presented in (\ref{derivation4}).  Furthermore, as we demonstrate in Section 4, the model computation is "so tractable" that we perform it in Microsoft Excel.

\begin{itemize}
\item[(5)]{The model should allow for the incorporation of person-level heterogeneity reflecting the fact that individual's
interarrival rates may vary across the inferential population.}
\end{itemize}

One nice feature of the model presented in (\ref{derivation4}) is that
introducing heterogeneity across units in their rate parameters, $\lambda_i$, is straightforward.
If, as is standard in many timing models, we assume that
the underlying rates are drawn from a Gamma distribution, $\lambda_{i} \sim {\rm Gamma}(r,\alpha)$,
we can increase the model flexibility at the expense of only one additional model parameter and also,
as per wish list item 1 when $c=1$, nest the negative binomial model.
Thus, when we combine our polynomial expansion Weibull count model in (\ref{derivation4}) with a Gamma heterogeneity distribution,
we now have a counting model that nests the Poisson and negative binomial.

In particular, the derivation of the Weibull gamma heterogeneous counting model, model [6] from Section \ref{modelframe},
is given as follows:

\vspace{-0.2in}

\begin{eqnarray}\label{heteromodel}
P(I(t) = i) & = & \int_0^{\infty} [\sum_{j=i}^{\infty}
\frac{(-1)^{j+i}(\lambda t^c)^j \alpha_j^i}{\Gamma(cj+1)}]
\Gamma(r,\alpha) d \lambda \\ \nonumber
 & = & \int_0^{\infty} [\sum_{j=i}^{\infty} \frac{(-1)^{j+i} (\lambda t^c)^j
\alpha_j^i}{\Gamma(cj+1)}] \frac{\alpha^r (\lambda )^{r-1}
e^{-\alpha \lambda }}{\Gamma(r)} d \lambda \\ \nonumber
 & = & [\sum_{j=i}^{\infty} \frac{(-1)^{j+i} (t^c)^j
\alpha_j^i}{\Gamma(cj+1)}] \int_0^{\infty} \lambda^j
\frac{\alpha^r (\lambda )^{r-1} e^{-\alpha \lambda }}{\Gamma(r)} d
\lambda \\ \nonumber
 & = & [\sum_{j=i}^{\infty} \frac{(-1)^{j+i} (t^c)^j
\alpha_j^i}{\Gamma(cj+1)}] \frac{\Gamma(r+j)}{\Gamma(r)
\alpha^j};
\end{eqnarray}

\noindent a solution that is quite "cute" as the solution to the integral in
(\ref{heteromodel}) is simply a weighted sum of the $j$-th moments of the gamma distribution around zero,
$\frac{\Gamma(r+j)}{\Gamma(r) \alpha^j}$, as $\lambda_{i}^{j}$ enters the polynomial approximated likelihood
in a linear way.  Hence, the conjugacy of the gamma distribution, and the polynomial approximated
likelihood is directly obtained.

\begin{itemize}
\item[(6)]{Covariate effects should be incorporatable as many (if not most) data sets have explanatory variables.}
\end{itemize}

Now that we have the closed-form solution for the counting model
whose underlying interarrival process is Weibull, and that
allows for heterogeneity in the underlying rates, we extend the model
to allow for the inclusion of covariates, i.e. models [5] and  [7] from Section \ref{modelframe},
by modifying the hazard function in a proportional hazards way.
In particular, we define the Weibull regression model, without heterogeneity, as

\vspace{-0.2in}

\begin{eqnarray}\label{model5}
P(I(t) = i) & = & \sum_{j=i}^{\infty} \frac{(-1)^{j+i}
(\lambda e^{x^{\prime}_{i} \beta} t^c)^j \alpha_j^i}{\Gamma(cj+1)} \\ \nonumber
 & = & (\sum_{j=i}^{\infty} \frac{(-1)^{j+i} (\lambda t^c)^j
\alpha_j^i}{\Gamma(cj+1)})(e^{x^{\prime}_{i} \beta})^j
\end{eqnarray}

\noindent  where $x^{\prime}_{i}$ denotes the
covariate vector for unit $i$ and $\beta$ a set of covariate slopes.
In an analogous manner, we derive model [7], our most complex model which allows for Weibull interrival times,
covariate heterogeneity, and parameter heterogeneity and is given by:

\begin{eqnarray}
\sum_{j=i}^{\infty} \frac{(-1)^{j+i} (t^c)^j
\alpha_j^i}{\Gamma(cj+1)}] \frac{\Gamma(r+j)}{\Gamma(r)
\alpha^j}(e^{x^{\prime}_{i} \beta})^j.
\end{eqnarray}

\noindent after integrating over $\lambda_{i} \sim {\rm Gamma}(r,\alpha)$.  We next describe an application of these models to a data set initially described and analyzed in Winkelmann (1995).

\section{Testing and Results}
\label{empirical}

Besides the derivation of the Weibull counting model, with and without covariates and with and without heterogeneity,
an additional goal of this research was to provide an empirical demonstration of our model with two aspects in mind.
First, that the polynomial expansion and conjugate prior derived here, which then allows for a closed-form solution
has computational advantages that should not be trivialized.  Remarkably enough, the computational
approach for our class of models, including the computation of bootstrap standard errors (Efron, 1982), was conducted
entirely in Microsoft Excel, an aspect we believe makes our approach widely accessible. The spreadsheets that were utilized
are readily available upon request.

Specifically, to compute the standard errors of coefficients under the series of
models, we utilized a bootstrap procedure in which
30 replicate data sets for each model were generated by sampling
individual respondents from the original data set with
replacement. The results reported for the standard errors are the
standard deviation of the coefficients across those samples.  We
note that for our model, the bootstrapping procedure can be
implemented by using a weighted likelihood approach where each
observations weight in the likelihood is the number of times in
which it randomly appears in the replicate sample; a procedure straightforwardly
implemented within Excel.  This equivalence of using a weighted likelihood approach
to compute bootstrap standard errors we believe is not specific to this model, can be utilized in a large number of research domains, and hence can be applied in software packages that contain just random number generation and function maximizer (e.g. Microsoft Excel solver) capabilities.

Secondly, one research
question we hoped to investigate was whether a more ``realistic'' (flexible) timing model (e.g. Weibull versus
Exponential) might alleviate the need for heterogeneity -- whether brought in through the underlying rates or via
covariates.  Thus, as we fit a sequence of models with increasing complexity (Poisson, Poisson with covariates, negative binomial, negative binomial with covariates, Weibull, Weibull with covariates,
Weibull with gamma heterogeneity, and Weibull with gamma heterogeneity and with covariates as
described in Section \ref{modelframe}), but differing in the source of that complexity, we focus on which aspects of the model are doing the ``heavy
lifting''.  Therefore, if in fact we find that a more rich underlying kernel timing model can provide an adequate fit as compared
to a model suggesting it is heterogeneity, this is important from a scientific perspective as these two "versions of the world" have very different underlying process models; and, it should be recognized that both are
plausible given the particular data and model.

In particular, we apply our series of models to a dataset initially (and more fully) described
in Winkelmann (1995)\footnote{We thank and acknowledge Professor Winkelmann for providing us this
data and helping us interpret his findings.} which contains as a dependent variable the number of children (a count) born
to a random sample of females.  In addition, a number of explanatory variables at the
subject level are available such as the female's general eduction (measured as the number
of years of school), a series of dummy variables for post-secondary education, either vocational training
or university, nationality (German or not), rural or urban dwelling, religious denomination (Catholic, Protestant, and Muslim,
with other or none as reference group), and continuous variables year of birth and age at marriage.

This data set was chosen for a number of reasons.  First, the paper by Winkelmann (1995) acted as a motivation and provided
a call for this research; hence utilizing the identical data set made sense.  Secondly, for this data set, the variance
of the number of births is less than the mean (2.3 versus 2.4), thus this data set provided an opportunity to demonstrate
the flexibility of the Weibull family of counting models derived here to handle underdispersion.  And finally, as Winkelmann (1995)
already contained the results for the Poisson regression model (model [1] here) and the Gamma timing model derived in
Winkelmann (1995), we already had results with which we could both confirm that our model and computational approach was error-free,
and provide a fairly high benchmark (the gamma timing model) to which we could compare the Weibull.

Tables \ref{non-reg-results} and \ref{reg-results} below list the results of the non-regression models (without covariates)
and the regression models.  We note that the log-likelihood values computed using our approach,
for both the Poisson model (LL = -2186.8) and Poisson model with
covariates (LL = -2101.8) are identical to that in Table 1 (p. 471) of Winkelmann (1995), thus verifying the accuracy of our polynomial expansion approach.  In addition, the
last column in Table \ref{reg-results}, the results of a Gamma counting model, is taken directly from Table 1 (p. 471)
from Winkelmann (1995).  We first describe our findings with
respect to the models without and then with covariates.

With respect to the non-regression models, the results indicate both some expected and informative
patterns.  First, the Weibull model has a higher log-likelihood
than the Poisson (which it must as it nests it) and the Poisson and NBD models are identical for
this data set as the underdispersion will drive the NBD heterogeneity to zero ($r$ and $\alpha$ are
extremely large) as the NBD gamma heterogeneity (if it existed) would overdisperse, not underdisperse, the fertility counts;
hence the NBD simplifies here (in essence) to the Poisson.
Similarly, the log-likelihood of the Weibull gamma
heterogeneity model is equal to that of the simple Weibull model as heterogeneity would add overdispersion to a data set that does not require it.
What this implies in combination is that once the richer Weibull counting model is utilized (which defeats both the Poisson and NBD), heterogeneity plays no role.

Therefore, a key observation to make here is that we now have the
ability to distinguish between dispersion in the individual-level
process, and dispersion due to heterogeneity.  That is, the
Weibull model allows for the detection of overdispersion and
underdispersion, but the gamma heterogeneity helps detect
dispersion in a totally different way, through the rate parameter.
If the underlying data set were instead overdispersed and a
Weibull gamma heterogeneity model were run on the data, one could
ask whether the dispersion effects were coming from the Weibull
model itself or from the dispersion due to gamma heterogeneity.
We leave this for future research and is discussed in Section 5.

Notice finally that the value of $c$ for the Weibull and Weibull
gamma heterogeneity models are both 1.116, slightly more than two
standard errors above 1 which is consistent with our result that
when $c$ is greater than 1, the Weibull counting model's variance is
less than the mean -- underdispersion.

\begin{center}
{\bf Insert Table \ref{non-reg-results} Here}
\end{center}

With respect to the regression models, we first describe some important
aggregate findings.  First, we note that the Weibull
regression model has a higher log likelihood than the Gamma count model
of Winkelmann (1995), albeit they are quite close. Moreover,
adding gamma heterogeneity to the Weibull regression adds little to the model;
again explained by the fact that the
dataset is underdispersed.  Similarly, adding gamma heterogeneity to the Poisson
regression model (the NBD regression) also does not raise the log-likelihood.
Notice finally that the values of $c$ for the Weibull regression and Weibull regression
with gamma heterogeneity models are 1.254 and 1.230, respectively, both significantly
greater than 1 (also indicated by the respective log likelihoods being sigificantly higher
than their Poisson and NBD regression counterparts) indicating underdispersion.

In terms of detailed findings for the covariates themselves, a number of ``verifying'' results emerge.
The coefficients of all variables are identical in sign as those in Winkelmann (1995), are extremely
stable across the class of models, and have essentially identical t-statistics (available upon request) such
that the variables that are significant coincide in both sets of
models\footnote{The year of birth and age of marriage variables were centered in Winkelmann, hence the difference in size of the coefficients.
However, the Poisson regression models as indicated by the log likelihoods are the same.}.  We expected the slight differences
in t-statistics that occurred due to the bootstrap standard errors utilized here and the
asymptotic ones from Winkelmann (1995).

Finally, in terms of the inferences from the coefficients themselves,
we focus on the results from the best fitting model, the Weibull Regression with Gamma heterogeneity.
As found in Winkelmann (1995),  non-Germans, people without vocational training, Catholics, Muslims,
and persons who get married at an early age all have significantly more children.  Maybe somewhat surprisingly,
rural versus urban and years of schooling do not.

\begin{center}
{\bf Insert Table \ref{reg-results} Here}
\end{center}

\section{Conclusions}
\label{conclusions}

In this research, we have derived and provided an empirical demonstration for an entirely new
class of counting models derived from a Weibull interarrival time process.  The new model has
many nice features such as it's closed-form nature, computational simplicity that can be
done in widely distributed software, the ability to nest both the Poisson and NBD models, and
the ability to bring in both heterogeneity and covariates in a natural way.  So, one may ask,

\begin{center}
{\em ``Is this the end? Is the problem solved?''}
\end{center}

\noindent We believe in one sense yes and in other ways no.  In the sense that we have derived
a model with the properties that were called for by Winkelmann and others -- the answer is yes.
However, we believe that the next phase, and one that is equally important, is the application
of our model to additional data.  Then, and only then, might we be able to improve our
understanding of some underlying processes to answer the question:

\begin{center}
{\em ``Is it heterogeneity or is it just Weibull?''}
\end{center}

\noindent With this paper's tools in hand, we believe that this is now an obtainable goal.

\section*{References}
\renewcommand{\bibitem}{\par\hang\hskip-\parindent}

\noindent \bibitem Arbous A.G., Kerrich J.E. (1951), ``Accident
Statistics and the Concept of Accident-Proneness'', {\em
Biometrics}, 341-433.

\noindent \bibitem Barlow, R. E., Proschan F. (1965),
``Mathematical Theory of Reliability", {\em John Wiley and
Sons}, New York.

\noindent \bibitem Bradlow, E.T., Hardie, B.G.S., Fader, P.S.
(2002), ``Bayesian Inference for the Negative Binomial
Distribution Via Polynomial Expansions'', {\em Journal of
Computational and Graphical Statistics}, Vol. 11, No. 1,
189-201.

\noindent \bibitem Cox, D.R. (1972), ``Regression Models and
Life-tables'', {\em Journal of the Royal Statistical Society.
Series B}, Vol. 34, 187-220.

\noindent \bibitem Efron, B. (1982), ``Bootstrap Methods : Another
Look at the Jackknife", {\em Annals of Statistics}, Vol. 7,
1-26.

\noindent \bibitem Everson, P.J. and Bradlow, E.T. (2002),
``Bayesian Inference for the Beta-Binomial Distribution via
Polynomial Expansions'', {\em Journal of Computational and
Graphical Statistics}, Vol. 11, No. 1, 202-207.

\noindent \bibitem Feller, W. (1943), ``On a General Class of
'Contagious' Distributions'', {\em Annals of Mathematical
Statistics}, Vol. 14, pp. 389-400.

\noindent \bibitem Gourieroux, C., Visser, M. (1997), ``A Count
Data Model with Unobserved Heterogeneity'', {\em Journal of
Econometrics}, Vol. 79, 247-268.

\noindent \bibitem Gurland, J., Sethuraman, J. (1995), ``How
Pooling Failure Data May Reverse Increasing Failure Rates'', {\em
Journal of the American Statistical Association}, Vol. 90
1416-1423.

 \noindent \bibitem Cameron, A. C., Johansson, P.
(1997), ``Count Data Regression Using Series Expansion: With
Applications'', {\em Journal of Applied Econometrics}, May, Vol
12, No. 3, 203-223.

\noindent \bibitem King, G. (1989), ``Variance Specification in
Event Count Models: From Restrictive Assumptions to a Generalized
Estimator'', {\em American Journal of Political Science}, August,
Vol. 33, No. 3, 762-784.

\noindent \bibitem Lawless, J. F. (1987), ``Regression Methods for
Poisson Process Data'', {\em The Journal of the American
Statistical Association}, Vol. 82, 808-815.

\noindent \bibitem Miller, S.J., Bradlow, E.T., and Dayartna, K.
(2004), ``Closed-Form Bayesian Inferences for the Logit Model via
Polynomial Expansions'', working paper.

\noindent \bibitem Mullahy, J. (1986), ``Specification and Testing
of Some Modified Count Data Models'', {\em Journal of
Econometrics}, Vol. 33, 341-351.

\noindent \bibitem Trivedi, P. K., Cameron A. C. (1996),
``Applications of Count Data Models to Financial Data'', {\em
Chapter 12 in Handbook of Statistics Vol 14: Statistics in
Finance}, North Holland, 363-391.

\noindent \bibitem Trivedi, P. K., Deb. P. (1997), ``The Demand
for Health Care by the Elderly: A Finite Mixture Approach'', {\em
Journal of Applied Econometrics}, Vol 12, No. 3, 313-332.

\noindent \bibitem Winkelmann, R. (1995), ``Duration Dependence
and Dispersion in Count-Data Models'', {\em Journal of Business
and Economic Statistics}, October, Vol 13, No. 4, 467-474.

\noindent \bibitem Winkelmann, R. (2003), ``Econometric Analysis
of Count Data'', Fourth Edition, Heidelberg, New York:
Springer''.

\newpage

\end{document}